\documentclass[aps,nofootinbib]{revtex4}
\usepackage{amsfonts}
\usepackage{mathrsfs}
\usepackage[dvipsnames,usenames]{color}
\usepackage{graphicx}
\usepackage{epstopdf}
\usepackage{ulem}

\def\rsim{\mathrel{\raise3pt\hbox to 8pt{\raise -6pt\hbox{$\sim$}\hss{$>$}}}}

\def\lsim{\mathrel{\raise3pt\hbox to 8pt{\raise -6pt\hbox{$\sim$}\hss{$<$}}}}
\def\rsim{\mathrel{\raise3pt\hbox to 8pt{\raise -6pt\hbox{$\sim$}\hss{$>$}}}}

\newcommand{\bh}{\mbox{\boldmath $h$}}

\newskip\humongous \humongous=0pt plus 1000pt minus 1000pt

\newif\ifdtup

\input colordvi

\begin{document}

\begin{widetext} 

\title{Ubiquity of Benford's law and emergence of the reciprocal distribution} 
\author{J.\ L.\ Friar$^1$\footnote{Electronic address: friar@lanl.gov}, 
T.\ Goldman$^1$\footnote{Electronic address: tjgoldman@post.harvard.edu}
 and J.\ P\'erez--Mercader$^2$\footnote{Electronic address:
jperezmercader@fas.harvard.edu}\footnote{Corresponding Author}
}
\affiliation{
    $^1$Theoretical Division, Los Alamos National Laboratory, Los Alamos, NM 87545     \\
$^2$
Department of Earth and Planetary Sciences,
    Harvard University, 
    Cambridge, MA 02138 \\
     and \\
    Santa Fe Institute, 
    1399 Hyde Park Road, 
    Santa Fe, NM 87501
}

\date{{\bf \today}}

\vspace{0.5cm}

\begin{abstract} 

\vspace{.5cm}

We apply the Law of Total Probability 
to the construction of scale-invariant probability distribution functions (pdf's), 
and require that probability measures be dimensionless and unitless under a continuous change of scales. If the scale-change distribution function is scale invariant then the constructed distribution will also be scale invariant.  Repeated application of this construction on an arbitrary set of (normalizable) pdf's results again in scale-invariant distributions. The invariant function of this procedure is  given uniquely by the reciprocal distribution, suggesting a kind of universality. 
We separately demonstrate that the reciprocal distribution results uniquely from requiring maximum entropy for size-class distributions with uniform bin sizes. 

\end{abstract} 

\maketitle

\vspace{.125in}


\end{widetext}

 
\pagebreak


\section*{INTRODUCTION}

\color{Black}
In 1881~\cite{newcomb} the astronomer and mathematician Simon Newcomb observed that the front pages of tables of logarithms were more worn than later pages. In other words mantissas corresponding to quantities that had a smaller first digit were more common than for quantities with a larger first digit. He argued that the distribution of ``typical'' mantissas was therefore logarithmic. The physicist Frank Benford~\cite{benford} rediscovered this in 1938 and provided more detail, for which his name is now associated with this phenomenon.

By now it is well documented that the frequency of first digits $D$ in the values of quantities randomly drawn from an ``arbitrary'' sample follows Benford's Law of Significant Digits, namely,

\begin{equation}
B_b (D) = \frac{\ln (1+D)- \ln(D) }{\ln (b)}  = \int_D^{1+D} \frac{dx}{x \cdot \ln(b)}\; ,
\label{1}
\end{equation}

\noindent
where $b$ is the arbitrary base for the logarithms and is commonly taken to be 10. We note that the probability of first digit 1 for base 10 is $\log_{10} (2) \cong .30$, far exceeding that for a uniform distribution of digits. The rightmost expression in Eqn.~(\ref{1}) expresses Newcomb's and Benford's logarithmic distribution as the cumulative distribution function (cdf) based on the {\bf reciprocal} probability distribution function (pdf), which has been normalized to 1. The pdf that underlies Benford's Law is therefore the reciprocal distribution, $r(x)\equiv c/x$, with normalization constant  $c=1/\ln b$ when the random variable $x$ ranges between $1/b$ and $1$. We note that Eqn.~(\ref{1}) is base invariant ({\it i.e.}, invariant under a common change in the base of the various logarithms) and that the reciprocal pdf is scale invariant (a function $f(x)$ is said to be scale invariant if $f(\lambda x)= {\lambda}^p \cdot f(x)$ for any $p \in {\mathbb C}$). In this work we will concentrate on the emergence of the reciprocal distribution under a variety of conditions. The invariant (or fixed-point) function of an iterative procedure applied to distribution functions that are invariant under a continuous change of scales will be shown to be the reciprocal distribution. Additionally, requiring maximum entropy for size-class distributions with uniformly distributed bin sizes leads to the same function.

Very relevant to the discussion above is T. P. Hill's proof in 1995~\cite{hill1,hill2,hill3,hill4} that random samples chosen from random probability distributions  are collectively described by the reciprocal distribution, which is the pdf for the logarithmic or Benford distribution. In Hill's words: ``If distributions are selected at random (in any ``unbiased'' way) and random samples are then taken from each of these distributions the significant digits of the combined sample will converge to the logarithmic (Benford) distribution.'' Because of this, the latter has been appropriately characterized as ``the distribution of distributions,'' as Hill's theorem is in some sense the obverse (counterpart) of the Central Limit Theorem for probability  {\bf distributions} with large numbers of samples.

Benford's Law has been found to hold in an extraordinary number and variety of phenomena in areas as diverse as physics~\cite{nuclear,atomic,astro,geo,statmech,ariel}, genomics~\cite{gref}, engineering~\cite{eref} and among many others, forensic accounting~\cite{nigrini}. Recently the number of examples where it applies has been expanding rather rapidly. 

In the 1960's the need for understanding the constraints imposed in computation by finite word length and its impact on round-off errors were behind the interest of many, including R. Hamming~\cite{hamming1,hamming2}, in Benford's law.

Importantly, Hamming argued that {\bf repeated} application of any of the four basic arithmetic operations (addition, subtraction, multiplication and division) to numbers leads to {\bf results} whose distribution of leading floating-point digits approaches the logarithmic (Benford) distribution. Hamming further argued that if any one arithmetic operation involves a quantity {\bf already} distributed according to the reciprocal distribution, $r(x)$, then the result of this and all subsequent operations will result in quantities whose pdf for the leading floating-point digits is the reciprocal distribution. Hamming called this property the ``{\bf persistence} of the reciprocal distribution'' although a better word might be {\bf contagiousness}, since contact with the reciprocal distribution at any point in a calculational chain modifies the remaining chain irrevocably. 

In this paper we use elementary methods to explore the connection between Benford's law, Hill's theorem and the ``contagiousness'' property of the reciprocal distribution.  We will demonstrate this by constructing a simple but comprehensive class of probability distributions that depends on a single random variable that is dimensionless and unitless under a continuous change of scales. This class depends on an underlying pdf that is arbitrary, and which can be sampled in a manner consistent with Hill's Theorem. We further generalize this into an iterative procedure whose invariant functions are shown uniquely to be the reciprocal distribution, and which demonstrate Hamming's ``contagiousness''. Uniqueness obtains because the arbitrary (or ``random'' in this sense) underlying pdf eliminates any particular solutions in the invariant functions and leaves only the general solution. Our procedure generalizes the work of Hamming\cite{hamming1}, and to the best of our knowledge is both new and useful. We show alternatively by invoking maximum entropy for a size-class distribution function that  the reciprocal distribution again obtains as the unique solution. We conclude by speculating on the universality and applications of these results, with particular emphasis on minimizing errors in computations of various types.

\section*{RESULTS}
\subsection*{Invariance under changes in units and the law of total probability}

In most scientific applications a stochastic variable $x$ is assigned to the random values of some physical quantity. This quantity carries either physical dimensions ({\it e.g.}, length or volume) or units (such as the number of base pairs in a genome). However, because it refers to probabilities, the probability measure $F(x)\cdot dx$ that characterizes $x$ must be dimensionless and unitless. 
 
Hence, in order to {remove} units or dimensions from the {measure} it is necessary to introduce a parameter that results both in a dimensionless and unitless stochastic variable, {\bf as well as in a bona fide} probability measure. Calling this parameter $\sigma$, for a specific value of $\sigma$ we can rescale the physical variable $x$ into a dimensionless and unitless random variable by just replacing $x$ with $z = x/\sigma$. (We also assume for simplicity that $x$ is positive semi-definite.) Then we can always introduce a normalizable function $g$ such that 
 
\begin{equation}
F (x) =  \frac{1}{\sigma}\; g\left(\frac{x}{\sigma}\right)
\label{F} \, ,
\end{equation}

\noindent
and that has the correct properties expected from a probability measure. 
In other words, we can use a parameter $\sigma$ to remove units or dimensions from the probability measure. 

Familiar examples of distributions of the type $g$ are the uniform distribution, $g_u (z) = \theta (1-z)$, the Gaussian distribution, $g_G (z) =\frac{2}{\sqrt{\pi}}\, \exp{(-z^2)}$, and the exponential distribution, $g_e (z) = \exp{(-z)}$, all of which satisfy the normalization condition: $\int_0^{\infty} d z \; g(z) = 1$. Heaviside step functions can be used for those cases where $g (z)$ is only non-vanishing in an interval, such as $z = [a,b]$,  as was done above for the uniform distribution. 
   
But the units chosen to measure $x$ are, of course, arbitrary. For example, if $x$ is a length, the units could be meter, millimeter, Angstrom, or even fathom, furlong, league, etc. In other words, the choice of units is itself arbitrary~\cite{pinkham} and we can think of $\sigma$ as a random variable with a distribution function $h(\sigma)$. 
Thus the problem we must study involves the combination of two stochastic variables. 
We can conveniently remove the scale and avoid the issue 
of units by using the Law of Total Probability~\cite{problaw} to combine 
the distribution $g$ with a distribution of scale choices to 
produce a distribution $G(x)$:
\begin{equation}
G(x)= \int_0^{\infty} d \sigma \; \frac{g(x\,|\,\sigma)}{\sigma}\, h(\sigma) \, ,
\label{LTP}
\end{equation}

\noindent
where now $G(x)$ and $h(\sigma)$ are interpreted as the {\bf marginal} probabilities for events $x$ and $\sigma$, and $g(x\,|\,\sigma)$ represents the conditional probability for $x$ given $\sigma$. This well known law captures the intuitively clear statement that the probability that event $x$ occurs is determined by summing the product of the probabilities for any of its antecedents $\sigma$ to happen, times the conditional probability that $x$ happens, {\bf given} that $\sigma$ has already occurred. Convergence of the integral for small values of $\sigma$  is not a problem for $x \neq 0$ if $g (z)$ vanishes sufficiently rapidly for large $z$. Normalizability of $g$ is sufficient for our purposes. The probability distribution in Eqn.~(\ref{LTP}) is fairly general and will be our template for studying the conditions underlying the emergence of the reciprocal distribution.

\subsection*{The Law of Total Probability and its recursive application}

Let us consider a $g(x|\sigma)$ that is invariant under changes in dimensions or units. That is, let us assume that 

\begin{equation}
g(x|\sigma) \equiv g\left(x/\sigma\right) ,
\label{interpretation}
\end{equation}
with a concomitant interpretation for $g(x/\sigma)$ in the terms described in the preceding paragraph (N.B. the difference between``$|$'' and ``$/$''). Changing the integration variable to $z\equiv x/\sigma$ in Eqn.~(\ref{interpretation}) leads to the convenient form

\begin{equation}
G(x)= \int_0^{\infty} \frac{d z}{z} \; g(z) \, h(x/z) \, .
\label{var}
\end{equation}

It is important to note a property of Eqn.~(\ref{var}) that is a consequence of its structure: the function $G(x)$ has an {\bf exceptional form} if $h(\sigma)$ is a scale-invariant (and power--law) function.
A scale-invariant function $h(x/z)$ must be a power of its argument, or $h(x/z) \propto (x/z)^{-s}$ for a power--law. Ignoring the (for now) irrelevant proportionality constant, we then have

\begin{equation}
G(x)= \frac{1}{x^s} \, \int_0^{\infty} d z \, z^{s-1} \; g(z)
\label{scale}
\end{equation}
for $h(\sigma) = 1/\sigma^s$. We note that the integral $\int_0^{\infty} d z \, z^{s-1} \; g(z)= {\cal{M}}_s (g)$ is a constant and the Mellin transform~\cite{inttrans} of the function $g(z)$. This allows one to rewrite Eqn.~(\ref{scale}) in the more compact form

\begin{equation}
G(x)= \frac{1}{x^s} \,{\cal{M}}_s (g) \, .
\label{firstrepeatedintegral}
\end{equation}
We therefore conclude that any scale-invariant $h(\sigma)$ in the integral in Eqn.~(\ref{var}) {\bf replicates itself} in $G(x)$ (up to an overall constant factor). Furthermore, this constant is equal to one if $h$ is the reciprocal distribution, since {\bf by definition} the Mellin transform in Eqn.~(\ref{firstrepeatedintegral}) equals unity for $s=1$  if $g(z)$ is normalizable.

\subsection*{Iterating the Law of Total Probability and the invariant function of the iteration}

With the above result in hand we are ready to tackle the following important question: what would be the result of applying the Law of Total Probability, as written in Eqn.~(\ref{LTP}), to a repeated and independent combination of random quantities if $h$ is scale invariant ({\it i.e.}, a power--law with exponent $s > 0$)? 

More specifically, suppose that we have $n$ random variables distributed according to the (in principle {\bf different}) distributions $g_1(z), g_2(z), \cdots g_n(z)$, and that $h$ is scale invariant. Defining the integral in Eqn.~(\ref{LTP}) or (\ref{var}) to be an integral transform operator, ${\cal{C}}$, acting on $h(\sigma)$, we can then operate ${\cal{C}}$ repeatedly on $G$ a total of $n$ times with (in principle) $n$ {\bf different} distributions to produce the $n$-th iterate. 

Denoting the result of the above operations by $G_n(x_n)$, we can then write it mathematically as

\begin{equation}
G_n (x_n) = \underbrace{C\cdot C \cdot \cdots C}_{n\ \mbox{\scriptsize times}} \{\bh\} \equiv {\cal{C}}^{\, n} \{\bh\} =  \frac{1}{x_n^s} \left [\prod_{i=1}^n {\cal{M}}_s (g_i)\right ] \, ,
\label{Cn}
\end{equation}

\noindent
where the constant in the large bracket is unity for $s=1$ {\bf and} if each $g$ is normalizable, as assumed. Equation~(\ref{Cn}) follows immediately from Eqn.~(\ref{scale}) in a natural way:
each succeeding application of ${\cal{C}}$ {\bf regenerates} the function $h$ and thus reproduces the previous application, except for an overall constant. This is of course the ``contagiousness'' or ``persistence'' property
noted by Hamming, which is the inevitable (unavoidable) result of using a scale-invariant prior function $h$ in Eqn.~(\ref{var}). (We note in passing that Hamming referred to scale-invariant distributions, but actually treated and discussed only the reciprocal distribution.)

We can extend and unify our discussion by examining the fixed-point functions, or more precisely the {\bf invariant functions}, of the iteration procedure in Eqn.~(\ref{Cn}). This is done by replacing $G_n$ with $h$, and results in
\begin{equation}
\bh = {\cal{C}}^{\, n} \{\bh\} 
\label{fixed}
\end{equation}
for {\bf any} $n$.  Thus the invariant function of the iterative procedure introduced in Eqn.~(\ref{Cn}) is the reciprocal distribution, since the  bracketed constant in that equation with multiple arbitrary distributions (and therefore normalizable) $g_n$ will only equal unity for $s=1$. 

This result should not come as a complete surprise: an invariant function cannot depend on the details of the arbitrary $g_n$. The scale $\sigma$ in the $g_n$ will couple to any scale in $h(\sigma)$ to produce arbitrary results unless $h$ is scale invariant, which is easy to demonstrate using a variety of simple distributions. Furthermore, the reciprocal distribution can be shown to be the unique solution when $g$ is the uniform distribution. Invoking a set of arbitrary pdf's ({\it i.e.}, the $g_n$) is similar in spirit to Hill's ``random distributions.'' In our case it rules out any {\bf particular} solutions to Eqn.~(\ref{fixed}), which will be different for each choice of $g$, but allows the scale--invariant {\bf general} solutions.

Consequently, {\bf the ``persistence'' or ``contagiousness'' of the reciprocal distribution is a property of the invariant-function solutions of the iterative procedure introduced in Eqn.~(\ref{Cn}).
} 

The preceding narrative was predicated on developing an understanding of the ``persistence'' of the reciprocal distribution. Nevertheless, we emphasize that the behavior of Eqn.~(\ref{fixed}) for $n=1$ is all that is needed in order to specify the reciprocal distribution as the {\bf unique} solution of
\begin{equation}
h(x)= \int_0^{\infty} \frac{d \sigma}{\sigma} g_i\left(\frac{x}{\sigma}\right)\, h(\sigma) \, ,
\label{none}
\end{equation}
where $g_i$ is any member of the set of arbitrary normalizable pdf's $\{g_1, g_2, \cdots\}$. The {\bf general} solution is clearly the reciprocal distribution, while any {\bf particular} solution will be determined by the intrinsic properties of a particular pdf, and cannot be a solution for all of them (or even a few of them).

\section*{DISCUSSION}
\subsection*{Selected applications to computing and to minimal truncation errors}

Scale invariance (which only restricts $h(x)$ to a general power law: $x^{-s}$) is thus seen to be a {\bf necessary} condition for the emergence of the reciprocal pdf (uniquely  $s = 1$) and its cumulative distribution function  that leads to Benford's Law of Significant Digits. 
 A {\bf sufficient} condition is that the reciprocal pdf emerge when in contact with arbitrary members of the set of normalizable pdf's. Alternatively, requiring an invariant-function solution to the ``contagion'' process {\bf leads} immediately and uniquely to the reciprocal distribution.

\subsection*{The reciprocal distribution and its contagiousness are  ``universal'' consequences of the repeated application of the Law of Total Probability}

We have established above that when combining data distributed according to a variety of pdf's into a common pdf for the result, the presence of scale invariance in at least one of the distributions for the data being combined leads to a scale-invariant common distribution. Moreover, if one of the distributions is the reciprocal, then all subsequent combined data is distributed according to the reciprocal distribution. It is like a mathematical paraphrasing of the popular 17th century English proverb  that ``Once a poor, always a poor'' into ``Once Reciprocal/Benford, always Reciprocal/Benford''.

The relationship of the reciprocal distribution to the invariant functions of the iterated Law of Total Probability was derived above.
We wish to suggest that this relationship is behind the abundant number and the wide spectrum of phenomena where Benford's Law of Significant Digits has been found to apply. Given the nature of these tightly constrained iterations, one can also envision how  the reciprocal distribution and the associated Benford's Law of Significant Digits could be considered {\bf universal} in a sense akin~\cite{mg-mjp-m} to the one used in condensed matter theory and for critical phenomena. We note that invariant-function solutions to iterative procedures are both extremely common and quite important in fields as diverse as chaotic dynamics~\cite{mitch}, theoretical ecology~\cite{may}, and many others.

In some sense, this result emerges as a consequence of the fact that the reciprocal distribution is the invariant function for iterations of the Law of Total Probability applied to algorithmic combination of physical data. Of course, it follows from the preceding analysis, that this also will be the case if one were to combine any data together with data that is purely numerical and classified according to first digits. 

\subsection*{Size - Classes and Maximum Entropy}

In addition to the properties just discussed, the reciprocal distribution has a remarkable property that impacts data transmission for applications in communication theory.

It follows from the functional form of the reciprocal distribution that the mean of reciprocally distributed random variables is a constant over any {\bf uniform interval}. If we could associate this mean with a probability distribution, then by grouping Benford--distributed data into uniformly distributed packages (corresponding to this average value) we could communicate those packages at the maximum information rate~\cite{shannon}.

We can formally implement the words above by considering the notion of {\bf size-classes}~\cite{ciofalo}. To that end we introduce the following construction for an arbitrary pdf, $p(x)$: 
\begin{equation}
\Delta\Phi[G[j]]  \equiv   \int_{G[j-1]}^{G[j]} dx \; x \; p(x) 
			 \equiv   \Phi[G[j]] - \Phi[G[j-1]] \, ,
\label{sizeclass}
\end{equation}
where $G[j]$ and $G[j-1]$ denote the upper and lower values of the sizes contained within the size-class indexed by $j$.  Since $p(x)$ is the pdf for $x$ taking values between $x_{\rm min}$ and $x_{\rm max}$,  we must have
\begin{equation}
\int_{x_{\rm min}}^{x_{\rm max}} dx \; p(x) = 1 \, .
\label{norm}
\end{equation}
The mean of $x$ with respect to $p(x)$ is then given by
\begin{equation}
\langle x \rangle = \int_{x_{\rm min}}^{x_{\rm max}} dx \cdot x \cdot p(x)  \, .
\label{xpectval}
\end{equation}

The probability that the values of the variable $x$ fall in the class $j$ ({\it i.e.}, between $G[j]$ and $G[j-1]$) is determined by  a new pdf, $p^{*}(j)$:
\begin{equation}
p^{*}(j ) =  \int_{G[j-1]}^{G[j] }dx \frac{x}{\langle x \rangle} p(x) = \frac{\Delta\Phi[G[j] ] }{\langle x \rangle} \, ,
\label{unifpdf}
\end{equation}
with the constraint from Eqns.~(\ref{norm}) and (\ref{xpectval}) that
\begin{equation}
\sum_{j=1}^{j_{\rm max}} p^{*}(j)  = 1 \, ,
\label{norm*}
\end{equation}
where $j_{\rm max}$ is the number of size-classes  into which the interval $[x_{\rm min}, x_{\rm max}]$ is partitioned. Thus for any pdf, size-classes and an associated pdf can be introduced. 

It is straightforward to show that the Benford distribution has the property that uniform size-classes are themselves uniformly distributed. If we {\bf require} $p^{*}(j)$ to be a constant (viz., $\alpha > 1$) that is independent of $j$, then
\begin{equation}
\Phi[G[j]] - \Phi[G[j-1]]  = \frac{1}{\alpha} \, ,
\label{p*jcnst}
\end{equation}
and the size-classes $G[j] - G[j-1]$ from Eqns.~(\ref{unifpdf}) and (\ref{p*jcnst}) that are found for $p(x) = a/x$ must satisfy 
\begin{equation}
 \frac{1}{\alpha}  =   \int_{G[j-1]}^{G[j]} dx\frac{x}{\langle x \rangle} p(x)  =   \frac{a}{\langle x \rangle} \left ( G[j] - G[j-1]  \right ) \, .
\label{sizeclassrln}
\end{equation}
That is, the classes that satisfy Eqn.~(\ref{sizeclassrln}) are uniformly distributed and therefore have maximum entropy; the information they contain can be transmitted at the maximum achievable rate: $H = +\ln(\alpha)$. 

For the reciprocal-distribution example we have
\begin{equation}
G[j] - G[j-1]  =   \frac{\langle x \rangle}{a \cdot \alpha} = \frac{x_{\rm max}-x_{\rm min}}{\alpha}  \equiv  \beta \, ,
\label{sizeclassval}
\end{equation}
where Eqn.~(\ref{xpectval}) has been used. The recursion relation  in Eqn.~(\ref{sizeclassval}) for the size-class boundary $G[j]$ is solved by 
\begin{equation}
G[j]  =   \beta j + r \, ,
\label{solution}
\end{equation}
where $r$ is an arbitrary constant. The integer quantity $j_{\rm max}$ in Eqn.(\ref{norm*}) can then be shown to equal $\alpha$ (which must also be integer).

Therefore, Benford-distributed data grouped into such size-classes will be communicated at the maximum achievable rate\footnote{ Note that for any distribution we can always introduce the notion of an {\bf $f$-class} (corresponding to a function $f$) by constructing the equivalent of $p^{*}(j)$ in Eqn.~(\ref{unifpdf}). All one needs to do is replace $\frac{x}{ \langle x \rangle} p(x)$ in Eqn.~(\ref{unifpdf}) by $\frac{f(x)}{\langle f(x) \rangle} p(x)$, where $f(x)$ is a function of the stochastic variable $x$ and represents some physical variable according to which we wish to sort the system into classes. If we require that the resulting $p^{*}(j;f)$ be constant, then the transmission of these classes will take place at the maximum rate. It is particularly interesting that for the Zipf distribution, $p(x) \propto 1/x^2$, the ``distortion-classes" that result from choosing $f(x) \propto x^2$ achieve maximum entropy in the same way as the size-classes do for the Benford distribution.}. Combining this result with the contagion property of Benford leads to the conclusion that the contagiousness of the reciprocal distribution via the LTP implies that a grouping into {\bf size-classes} of stochastic variables, at least one of which is Benford distributed, has maximal entropy as long as the grouping is uniformly partitioned. These classes or groupings of the original data can then be transmitted at the maximum achievable rate.

\subsection*{Round-off error and the Reciprocal Distribution. Applications to computing and to evolutionary biology}

The above observation can be turned into a principle of design that can be applied if one is interested in achieving maximum precision (equivalent to minimizing resulting errors by always choosing the smallest intermediate errors) in calculational situations where errors are unavoidable, as is the case for round-off errors occurring in automatic digital computation. This is what was suggested by Hamming's result mentioned in the introduction to this paper.

In fact, the contagiousness property of the reciprocal and the associated preponderance of mantissas starting with smaller digits indicates that distributing quantities according to the reciprocal distribution will produce maximal reduction in round-off errors dictated by the necessary truncation of results of operations due to the fixed and finite word length of the machine on which the operation takes place. 

This contagiousness property of the reciprocal pdf, combined with its relationship to the Benford Law of Significant Digits, was then used by Hamming to argue that the mantissas of the round-off errors that result from the arithmetical combination of random quantities are always smaller than what would result if they were uniformly distributed. Hence the arithmetical combination of quantities of which at least one is reciprocally distributed leads to smaller errors than those resulting from the combination of all-uniformly distributed quantities.

This implies for round-off error that it ``pays'' to design the data to be reciprocally distributed, since when combined it leads to minimal errors. This leads to a maximum reduction of the unavoidable round-off errors inherent~\cite{jlf} to any calculation done with digital computers that, of course, are ultimately due to the fixed nature of their word length~\cite{vonneumann,bernstein}. Indeed this was discussed by Schatte~\cite{schatte} as the basis for a strategy designed to reduce the accumulation of errors in the operation of digital machines (cf. also~\cite{knuth}). 
We combine this property with the above result that {\bf maximum entropy} is achieved by grouping  Benford-distributed data into uniformly sized classes. {\bf This demonstrates that those groups will automatically  be  the fastest transmitted packets with the smallest possible errors in both their transmission and contents.}

In this context we note  that the fundamental operations in information handling by the genomes of all forms of life are controlled by the so-called cDNA fraction of the genome.  This fraction comprises the {\bf fundamental} genes in the living system, and  are contained in objects called Open Reading Frames (ORFs). For all living systems the ORFs are distributed according to a reciprocal distribution of the full genome size~\cite{gref}, not just of only the cDNA fraction. This suggests to us that Life, by means of the trial and error processes of evolution, may have ``discovered''  the most robust and lowest-error strategy for storing and transmitting information. 
The key ingredient for distributing quantities  with maximal fidelity  using the Law of Total Probability (equivalent to lowest possible error) is to incorporate the reciprocal distribution. 
Subsequently the processes depending on these {\bf fundamental} genes will be the most robust and will therefore have a competitive advantage as a base to persist, survive and not become extinct.


\section*{ACKNOWLEDGMENTS}

One of us (JP-M) would like to thank the everis Foundation and Repsol for generous support, and the Theoretical Division of Los Alamos National Laboratory for its hospitality.




\end{document}